\documentclass[preprint,12pt, a4paper]{elsarticle}

\usepackage[T1]{fontenc}

\usepackage{hyperref}
\usepackage{palatino}
\usepackage{xcolor}
\usepackage{listings}

\definecolor{mygreen}{rgb}{0,0.6,0}
\definecolor{mygray}{rgb}{0.5,0.5,0.5}
\definecolor{mymauve}{rgb}{0.58,0,0.82}
\definecolor{codegreen}{rgb}{0,0.6,0}
\definecolor{codegray}{rgb}{0.5,0.5,0.5}
\definecolor{codepurple}{rgb}{0.58,0,0.82}
\definecolor{backcolour}{rgb}{0.95,0.95,0.92}

\lstdefinestyle{somestyle}{
    backgroundcolor=\color{backcolour},   
    commentstyle=\color{codegreen},
    keywordstyle=\color{blue},
    numberstyle=\tiny\color{codegray},
    stringstyle=\color{codepurple},
    basicstyle=\ttfamily\footnotesize,
    breakatwhitespace=false,         
    breaklines=true,                 
    captionpos=b,                    
    keepspaces=true,                 
    numbers=left,                    
    numbersep=5pt,                  
    showspaces=false,                
    showstringspaces=false,
    showtabs=false,                  
    tabsize=2
}
 
\newcommand{\jpet}{\mbox{J-PET}} 

\usepackage[margin=2.5cm]{geometry}

\usepackage{amssymb}

\usepackage{lineno}
\usepackage{float}
\restylefloat{table}
\journal{SoftwareX}

\begin{document}

\begin{frontmatter}



\title{J-PET Framework: Software platform for PET tomography data reconstruction and analysis}


\author{W. Krzemien}
\address{High Energy Physics Division, National Centre for Nuclear Research, \\
 05-400 Otwock-\'Swierk, Poland \\
 wojciech.krzemien@ncbj.gov.pl}

\author{A. Gajos}
\author{K. Kacprzak}
\author{K. Rakoczy}
\author{G. Korcyl}
 \address{Faculty of Physics, Astronomy and Applied Computer Science \\Jagiellonian University\\
S. \L{}ojasiewicza 11, 30-348 Krak\'ow, Poland \\}

\begin{abstract}
\small
J-PET Framework is an open-source software platform for data analysis, written in C++ and based on the ROOT package. It provides
a common environment for implementation of reconstruction, calibration and filtering procedures, as well as for user-level analyses of Positron Emission Tomography data.
The library contains a set of building blocks that can be combined by users with even little programming experience, into chains of processing tasks through a convenient, simple and well-documented API.  
The generic input-output interface allows processing the data from various sources:  low-level data from the tomography acquisition system or from diagnostic setups such as digital oscilloscopes, as well as high-level tomography structures e.g. sinograms or a list of lines-of-response. Moreover, the environment can be interfaced with Monte Carlo simulation packages such as GEANT and GATE, which are commonly used in the medical scientific community.

\end{abstract}

\begin{keyword}
Positron Emission Tomography \sep PET  \sep data processing \sep software analysis framework 
\end{keyword}

\end{frontmatter}

\noindent \footnotesize{Code metadata} 
\noindent
\begin{table}[ht]
\footnotesize
\begin{tabular}{l l}
\hline
Current code version & 8.1 \\

Permanent link to the repository & \cite{framework-repo} \\ 

Legal Code License & Apache License 2.0 \\

Code versioning system used & git \\ 

Software code languages, tools  and services used & C++, Python, ROOT, Boost \\

Compilation requirements \& dependencies & Linux; g++ compiler supporting the c++14 standard;  \\  & CMake 3.1.5 or later; Boost libraries 1.64 or later, \\ & ROOT version 6.X \\

Developer documentation/manual & \cite{framework-manual}  \\

User and developer support, bug tracker & \cite{framework-redmine} \\

\hline
\end{tabular}

\end{table}




\section{Motivation and significance}
\label{motivation}


Positron Emission Tomography is one of the most popular methods for tomographic imaging used in nuclear medicine. 
In contrast to other techniques such as Computed Tomography that can detect anatomical changes, 
PET provides information about metabolic processes in the patient's body even at the cell level~\cite{dbailey}. 
This allows detection of pathological symptoms that usually precede the anatomical changes. 
PET tomography has a wide range of research and clinical applications e.g.\ it is commonly used for diagnosis of cancer, neurological disorders, heart diseases and many others~\cite{Slomka}.

Although the PET technique is well established for clinical usage, there are ongoing efforts in the scientific community that would overcome the limits of the commercial scanners  
and improve the quality of the image~\cite{Slomka,Borghi_2016,8049484,Karpetas,Cates_2016} or even enrich the available information by introducing new diagnostic methods.
Whole-body or total-body PET scanner projects~\cite{Badawi2019,Cherry2017,Cherry2018} propose tomographs that improve the sensitivity of the measurement in order to shorten the time of a scan or alternatively require a smaller radiation exposure for the patients~\cite{NATURE_EXPLORER}. 

The transformation of the data acquired by a PET scanner from the \textit{raw} binary level till the final patient image analysed by physicians is a complex, multi-stage process involving low- and high-level reconstruction algorithms.
The associated data handling and reconstruction is an especially hard task in case of the whole-body scanners due to large data volume~\cite{Zhang2017}.

The J-PET collaboration aims at providing a low-cost, modular, whole-body PET scanner based on detection of photon interactions in plastic scintillators~\cite{NIM2014,Szymon-Acta, PMB2018} with a view to its application in both medical diagnostics~\cite{PMB2018, moskal:pmb2019} and in proton therapy monitoring~\cite{rucinski2020}.
The \jpet{} prototype is a research device which not only demonstrates the new operating principle for its use in standard PET tomography but also explores new imaging modalities such as spatially-resolved determination of properties of positronium atoms produced in a patient's body~\cite{imaging_patent, Jasinska-Moskal2017, daria_epjc, NATURE}. 

The exploratory nature of the \jpet{} device results in its operation
with much more flexible data registration conditions than used in commercial PET solutions.
In order to allow for classical PET imaging without discrimination of signals, which may be used in the novel diagnostic methods, \jpet{} operates
in a trigger-less data acquisition mode~\cite{Korcyl-IEEE}, resulting in a volume of recorded data unprecedented in medical imaging technologies.

From the software point of view, development and testing of novel PET modalities and tomography methods become challenging as the standard approaches must be either extended or entirely replaced by new algorithms.
Moreover, at the prototyping stage, multiple elements of the detector, its geometrical setup and the data acquisition chain are subject to change and various reconstruction procedures may be tested in parallel.
The software framework used to analyze data from evolving prototypes and to implement and test new reconstruction algorithms must follow these changes dynamically.
At the same time, however, the need to efficiently process the data stream from trigger-less acquisition requires that the performance may not be compromised when asserting flexibility.

The \jpet{} Framework package has been developed as an answer to the aforementioned challenges, providing a dynamically adjustable environment for development and efficient implementation of new algorithms. 
The basic idea is to provide a set of generic building blocks, allowing a quick implementation of data processing chains to be used by analysts with even little programming experience through a convenient and simple API. 
The \jpet{} Framework is used for analysis of data recorded by the tomograph prototype from the level of raw data saved by its data acquisition system, through assembly of higher-level data structures representing the logic needed for reconstruction of the physical properties of electron-positron annihilation into photons, up to the level of medical image reconstruction and statistical analysis of the data.

Notably, the data acquisition system of \jpet{} is based on the TRB3 hardware platform~\cite{Traxler_2011, Neiser_2013} which is widely used by experimental setups both in the fields of medical imaging and particle physics experiments \cite{TRB3users}. Consequently, the \jpet{} Framework can be easily adopted for data analysis in other TRB3-based experiments. 

From the point of view of the full data reconstruction flow, the usage scope of the \jpet{}  Framework is different compared with the existing tomography image software packages such as  STIR~\cite{stir}, CASTOR~\cite{castor} or QETIR~\cite{qetir}, since 
it also allows to implement low-level reconstruction and calibration algorithms  which operate  before the formation of Line-of-Responses (LORs), while typical input data for image algorithms consists of higher-level structures such as sinograms or list of LORs. At the same time, \jpet{} environment provides tools for the implementation of typical image reconstruction algorithms and effectively such procedures e.g. Time-of-Flight Filtered-Backprojection, have been implemented within the Framework.
However, the aim of the \jpet{} Framework package is not 
to replace the existing image tomography toolkits, which offer well tested and proven solutions, but rather to provide a possibility for passing the transformed data to the external packages.

While an early version of the \jpet{} Framework is described in Ref.~\cite{Krzemien2015Framework}, 
this article is intended to present its architecture and functionality available in its current mature form,
which allows to extend the scope of its usage beyond the \jpet project. Therefore, we focus on the properties of the core \jpet{} Framework library~\cite{framework-repo} rather than on the particular \jpet-specific reconstruction algorithms developed using the framework which are available in a separate repository~\cite{examples-repo}.

\section{Software description}
\label{soft-desc}

The design of the \jpet{} Framework originated from the necessity of performing reconstruction and analysis of PET data from a prototype tomography scanner. It has become a more generalized environment for execution of tasks which could be adapted to multi-step analysis of various kinds of data. All the features naturally followed the implementation: managing input/output, incorporating palette of configurations, adapting data and parameter structures, user interface and task handling. 

The core of the J-PET Framework is
constituted by
a dynamic library that can be linked to user applications. The library provides tools for loading, analysing and saving transformed data as well as for implementation of transformation algorithms that can be further connected in chains and finally executed.

The \jpet{} environment can be used to e.g. develop a reconstruction chain for the real data collected by a PET scanner or to implement a calibration procedure, an image reconstruction method or any kind of a multi-step analysis.  Other typical applications consist of comparative studies of prototype PET scanners performance based on the Monte Carlo (MC) simulations. 

\subsection{User Application Programming Interface}
\label{api}

\begin{figure}[h]
    \centering
    \includegraphics[width=\textwidth]{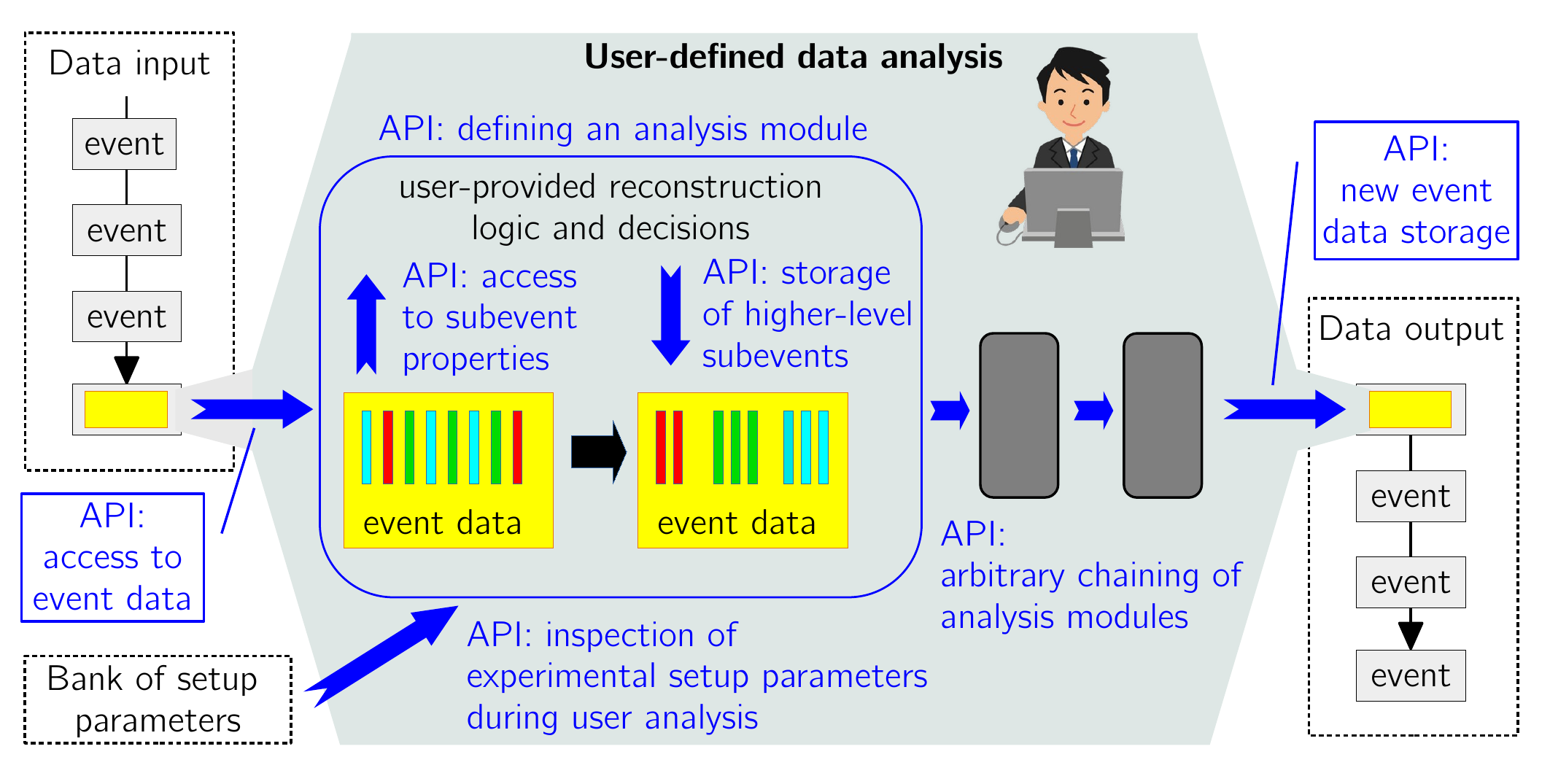}
    \caption{Scheme of the Application Programming Interface exposed to the user for creation of data reconstruction and analysis workflows. Elements of the workflow exposed to and modified by the user are contained in the middle gray region whereas the elements outside it are handled by the framework transparently to the user. Main functionalities of the API exposed to the user are marked with blue arrows and blue text.}
    \label{fig:api_scheme}
\end{figure}

The core library of the J-PET Framework provides the users with an Application Programming Interface (API) presented schematically in Figure~\ref{fig:api_scheme}.
The API is concentrated on giving the user access to data (structured as a stream of subsequent \textit{"events"}) read from various sources as well to parameters of the experimental setup (read from external configuration files) in order to combine
the event data and setup details in user-defined algorithms assembling higher-level
data structures and filtering them based on custom conditions (Listing~\ref{lst:event_finder} presents an simple example of such procedure).
The API further allows for grouping of such user-provided logic into analysis modules which can then be chained to constitute a complete analysis workflow (as demonstrated in Listing~\ref{lst:main}).

\subsection{Software Architecture}
\label{arch}

%
The library is written in C++ using object-oriented paradigm. The core components are implemented as classes with well-defined responsibilities e.g. computing task execution, input/output operations, logging, option parsing, option validation. Moreover, the package contains a set of classes representing physical entities e.g. part of the scanner or PET-specialized data structures such Line-of-Response, which form a language that can be used to express the domain-specific concepts (see more details in section~\ref{sec:param-data}).  
 
\begin{figure}[h]
\includegraphics[width=340pt,height=220pt]{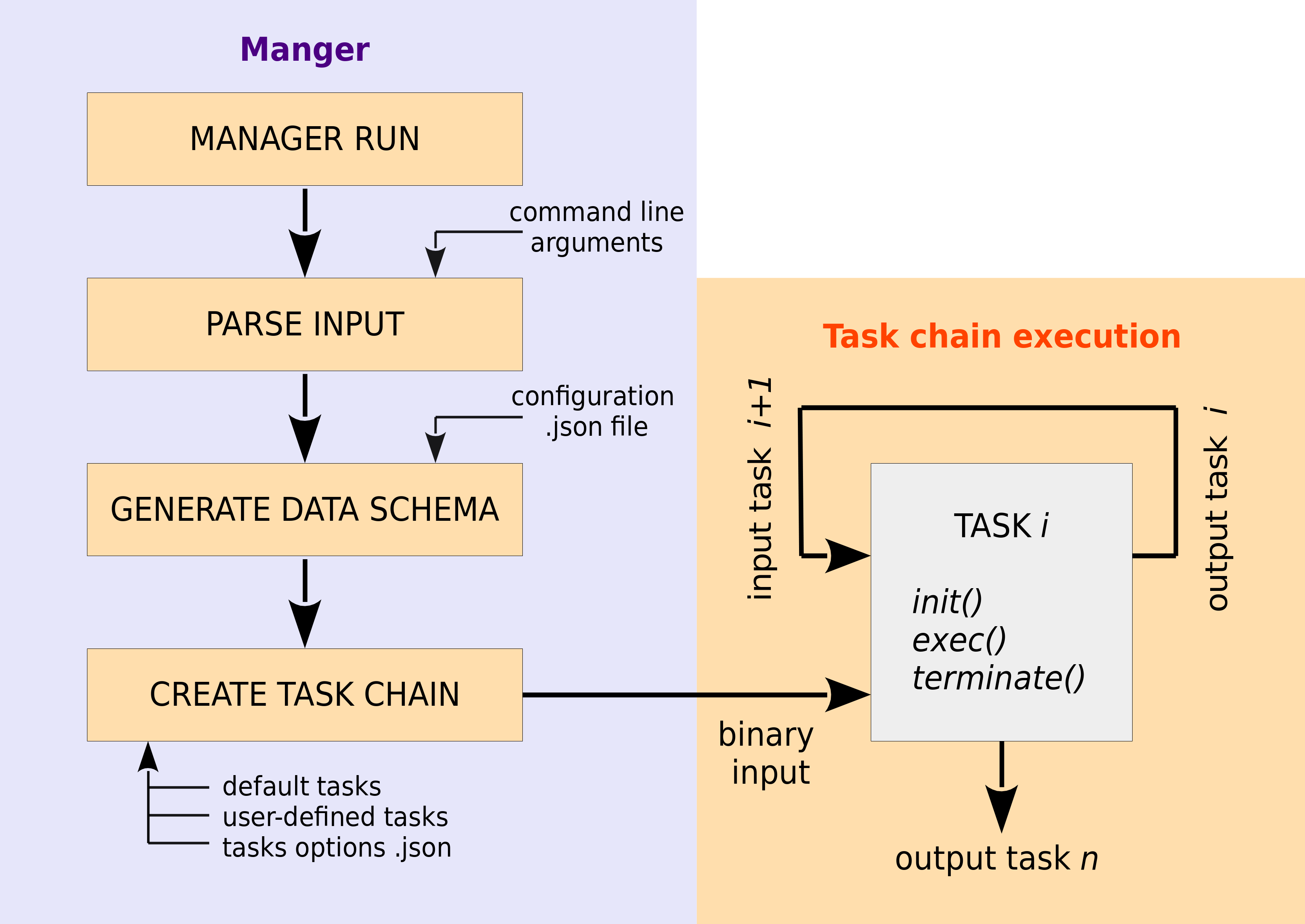}
\centering
\label{fig:manager}
\caption{Scheme of Frameworks \texttt{JPetManager} structure, showing order of initialization and execution of tasks.}
\end{figure}

The basic concept of the \jpet{} Framework is the decomposition of a data processing chain into a series of standardized modular blocks. Each module corresponds to a particular computing task, e.g.\ a reconstruction algorithm or a calibration procedure, with well-defined input and output. The processing chain is built by registration of chosen modules in the \texttt{JPetManager}, responsible for synchronization of the data flow between the modules (see Figure~\ref{fig:manager}). This approach
permits to quickly interchange modules and to create processing chains
for different experimental setups.

\subsection{Software Functionalities}\label{se:func}

\jpet{} Framework provides a set of functionalities that helps in rapid data reconstruction and analysis prototyping.
In this section, we list the most useful features and present selected usage examples. More applications can be found in the repository~\cite{examples-repo}.

\begin{figure}
    \centering
    \includegraphics[width=\textwidth]{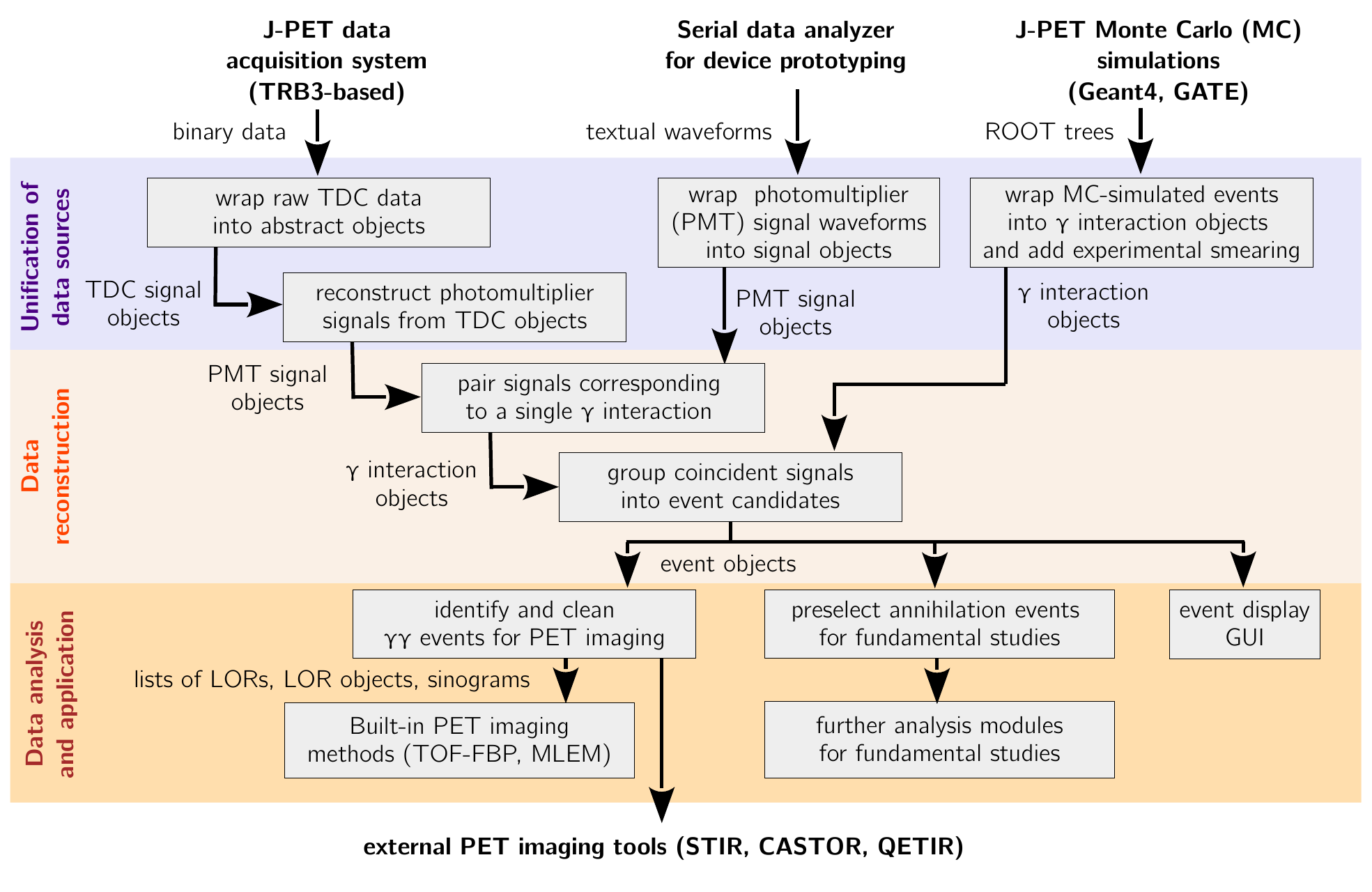}
    \caption{Scheme of the data processing paths realized with the J-PET Framework for different cases of analysis of data from the \jpet{} prototypes and the corresponding Monte Carlo simulations. Each gray rectangle represents a single module whereas arrows denote the flow of data represented as abstract objects.}
    \label{fig:large_scheme}
\end{figure}

\subsubsection{Handling of multiple data sources}\label{sec:inputs}

The Framework provides a generic input-output mechanism that through simple extensions allow for processing of data from various sources e.g. low-level data in a binary format from a tomographic data acquisition system, textual representations of complete photomultiplier (PMT) signal waveforms collected using a serial data analyzer at detector testing stages, as well as high-level tomography-specific structures e.g. sinograms or lines-of-response. Other extensions of the input interface feature using results of Monte Carlo simulations in place of data as described in section~\ref{sec:mc}.

Figure~\ref{fig:large_scheme} presents how inputs from various data sources are unified at higher analysis levels so that more abstract steps of reconstruction can act on data independently of their origin. PMT signals recorded with a serial data analyzer, for example, correspond to PMT signal representations already assembled from single Time-to-Digital Converter (TDC) signals in case of the TRB3-based data acquisition system and are thus injected to the analysis chain at the corresponding level, i.e.\ before a module pairing PMT signals from the same detection module to identify photon interactions.

\subsubsection{Input/Output mechanisms}\label{sec:io}

Besides the source-specific data formats handled by dedicated wrapper modules, the \jpet{} Framework relies on binary internally-compressed data format provided by the ROOT package, widely adopted in both particle physics and nuclear medicine research. The framework provides automatic handling of input and output files for standard data analysis modules, abstracting the actual storage away from the analysis or reconstruction logic. User code is only responsible for deciding whether an entry processed by a module should be preserved or discarded. Depending on the option chosen by the user, output from every analysis module is either saved to a separate file in the ROOT format or directly fed as input to the subsequent module in the chain. While the former is useful at the stages of testing the analysis, the latter approach allows to create a pipeline of analysis modules minimizing I/O load as the only output saved to disk is the one produced by the last module in the chain which typically corresponds to the most filtered data stream where the data volume is reduced by 1-2 orders of magnitude with respect to the raw input. This is particularly important when multiple analysis processes are operating on the same disk space, which is a common use case in data-driven parallel computing specific to particle physics and low-level PET tomography event filtering and reconstruction.

\subsubsection{Options}

The library provides multiple manners of loading optional information for any custom processing task. These collections of various parameters would be required for a successful reconstruction of PET data, giving i.e. descriptions of a experimental setup, measurement conditions, necessary calibrations or desired form of the output. The Framework provides the following interfaces for dynamic configuration:
\begin{itemize}
    \item command line options (e.g. input file, configuration files, progress display)
    \item \texttt{JSON} file with the description of experimental setup - parametrization of objects, that serve as data schema,
    \item \texttt{JSON} file with user-provided options -
      any custom settings to be used during execution of tasks, passed as named parameters of elementary C++ types
\end{itemize}

All the provided options are parsed and validated before execution of chain of tasks and
are accessible
during its processing.

\subsubsection{Data and parameter structures}\label{sec:param-data}

The library includes classes representing abstract entities
common for analysis of data from J-PET measurements.
Parameter Objects represent hardware parts of the detector,
along with their working parameters, in-setup placement and connections with other parts,
e.g.\
a single object per each plastic scintillator strip with its location in the detector or
a photomultiplier coupled to a given scintillator.
Data Objects are structures representing subsequent stages of reconstructed data --
from elementary ones containing only TDC time and data acquisition channel number, to a detailed reconstruction of a physical event or  a line-of-response. 

Data Objects refer to particular elements of the setup encapsulated in Parameter Objects where the physical signals have originated. Moreover, mapping of connections between such components imposes relations between Parameter Objects themselves. These relations are implemented using persistent object references (\textit{TRef}) provided by the ROOT libraries~\cite{root} which ensure $\mathcal{O}(1)$ lookup of corresponding elements as well as persistence of the relations across file storage.

On the user side, encapsulation of data and setup properties into abstract objects allows
for definition of reconstruction and analysis logic even by users without programming proficiency which is one of the objectives of the \jpet{} Framework. Listing~\ref{lst:event_finder} demonstrates the interplay between Data (\texttt{JPetHit} and \texttt{JPetEvent}) and Parameter Objects (\texttt{Scintillator} and \texttt{BarrelSlot}) in a simple task.

\subsubsection{Setup description}\label{sec:setup}

Since experimental setup and its conditions can change from one measurement to another, the set of parameters describing it must be generated dynamically. 
The library  provides dedicated tools to handle a setup representation in a form of a configuration \texttt{JSON} file. 
Based on its content, the Framework generates the collection of  Parameter Objects (see section~\ref{sec:param-data}) together with relations between them expressed in a standardized format.
Once this file is parsed, a Parameter Bank encapsulating the latter is embedded in all output data files to allow for their further stand-alone analysis.

\subsubsection{Processing control and logging}

Each execution instance of any application based on the Framework environment, generates a log file with a unique name. By default all input parameters and options are stored in the log file, moreover each task can produce custom messages with one of appropriate tags: \texttt{INFO}, \texttt{DEBUG}, \texttt{ERROR}. 

User Tasks classes can use tools for creating control histograms. All such objects are then automatically stored in the output file. An usage example can be found in code snippet \ref{lst:stats}.

\subsubsection{Compressed input files}
It is also possible to provide a raw data file in a compressed format; in that case a task is added by default, that simply decompresses the input file before any other procedures begin. Supported formats are: \texttt{xz}, \texttt{gz}, \texttt{bz2}, \texttt{zip}.

\subsubsection{Handling of binary data format}

The library can read raw data input files provided from the scanner data acquisition or from digital oscilloscope measurements. Binary data is transformed with dedicated tasks in the \texttt{ROOT} format, making it available for further processing by the consecutive tasks in the stream. 

\subsubsection{Iterative tasks}

The structure of task chain allows the implementation of iterative tasks schemes, in which a module can be executed in a loop till a given condition is fulfilled. The stopping condition can be based on desired number of consecutive iterations or on the return value of the function defined by the user. This functionality is especially useful for  optimization goals, i.e. refining detector calibration constants or estimation of event classification parameters. 

\subsubsection{Interfaces to Monte Carlo simulation packages}
\label{sec:mc}
Testing and debugging of data analysis modules is often supported by using Monte Carlo-simulated events in place of actual data. To this end, the \jpet{} Framework offers interfaces to two Monte Carlo simulation packages: the custom \jpet{} MC simulation software~\cite{mc-repo} based on the Geant4 toolkit~\cite{geant4} as well as the GATE package for simulation of PET and SPECT tomography~\cite{Gate_2004}.

MC-simulated events are wrapped into the same data structures as data so that analysis modules intended to process experimental data can be applied transparently to the simulation results. At the same time, all MC-specific event information is preserved and accessible on demand.

\subsubsection{Event Display}
\label{sec:event-display}
J-PET Event Display~\cite{event-display-repo} is a visualization tool based on the \jpet{} library. It can load files with the Framework data structures to visualize the reconstructed PET data in an event-by-event manner at different phases of the processing. Information on input geometry of the detector is provided by the same configuration files in the \texttt{JSON} format used for reconstruction of data described in section~\ref{sec:setup}. A usage example of the Event Display is shown in Fig.~\ref{fig:eventDisplay}.

\subsection{Development philosophy, testing and continuous integration}

The J-PET Framework developer community is trying to consistently adopt good coding rules and practices in the development routine to assure the quality of the software. In particular, any new code before being merged into official repository must be reviewed and accepted by at least one person not being the author. Moreover, it must pass a set of unit and integration tests defined for the platform. 
The contributors are strongly encouraged to add unit tests together with new classes and to format the code consistently using the clang-format tool.

The Continuous Integration process is integrated with the project Github repository. Any new pull request launches automatic set of tests based on the Travis~\cite{travis} and Jenkins~\cite{jenkins} services.
The  unit tests are operated by the Travis system, while larger integration tests, which typically require some input data, are run by a dedicated Jenkins server. Both services deliver a detailed report about possible failures.
The testing system is fully automatized on the servers and can be launched manually for local testing.

Issue and bug tracking is performed with the Redmine service which also serves also as a user support forum. The reference guide is automatically generated from the code using the Doxygen tool
and is available online~\cite{framework_doxygen}.
Additionally, an analysis user guide is provided in the repository and it is being updated with every new version of the Framework.

\subsection{Sample code snippets analysis}
\label{snippets}

Listing~\ref{lst:main} presents the instance of \texttt{JPetManager} registering user tasks to form a chain of procedures. With the following \texttt{useTask} method, the user is specifying the input and output data format of each tasks. In the example, the output of the first task serves as  input for the second one. The processing of all algorithms with the provided arguments is triggered by the \texttt{run} method. This simple construction allows to create an analysis from custom building blocks even for a user with little programming experience. 

Listing~\ref{lst:event_finder} presents a snippet of an analysis module identifying 2-photon coincidence events in a stream of single recorded photon interactions (referred to as \textit{hits}).
Listing~\ref{lst:stats} shows a basic usage of the statistics facilities for creation of histograms to be filled during data analysis.

\begin{lstlisting}[language=C++, label={lst:main}, caption=Exemplary main class of a program based on J-PET Framework library.]
#include <JPetManager/JPetManager.h>
#include "Task1.h"
#include "Task2.h"

using namespace std;

int main (int argc, const char * argv []) {
    JPetManager& manager = JPetManager::getManager();
    
    manager.registerTask<Task1>("Task1");
    manager.registerTask<Task2>("Task2");
    
    manager.useTask("Task1", "data.input", "data.type1");
    manager.useTask("Task2", "data.type1" , "data.type2");

    manager.run(argc, argv);
}
\end{lstlisting}

\begin{lstlisting}[
    language=C++, 
    label={lst:event_finder},
    caption=Exemplary naive procedure of finding 2-photon coincidence events demonstrating the ease of operations on the data structures provided by the Framework.
    ]
for(JPetHit& hit_1: gamma_hits){
  for(JPetHit& hit_2: gamma_hits){
    // find double coincidences within 5000 ps
    if(hit_2.getTime() - hit_1.getTime() < 5000.){ 
      // check if the two gamma interactions were recorded
      // in distinct scintillators of the setup
      if(hit_1.getScintillator() != hit_2.getScintillator()){
        // check if locations of the two detection modules
        // differ by more than 160 degrees in azimuthal angle
        if(fabs(hit_1.getBarrelSlot().getTheta() - 
                hit_2.getBarrelSlot().getTheta()) > 160.){
          // reconstruct e+e- -> 2gamma annihilation point
          TVector3 point =
            EventCategorizerTools::calculateAnnihilationPoint(hit_1,hit_2);
          // assemble an event containing the two hits
          JPetEvent event;
          event.addHit(hit_1);
          event.addHit(hit_2);
          event.setEventType(JPetEventType::k2Gamma);
          // automatically store the event in the output file
          // or pass on to the next analysis module
          fOutputEvents->add<JPetEvent>(event);
        }
      }
    }   
  }
}
\end{lstlisting}

\begin{lstlisting}[language=C++, label={lst:stats}, caption={Example of using tools for creating filling histograms, that are stored in output files.}]
// Creating histogram with JPetStatistics class
getStatistics().createHistogram(
  new TH1F("hit_z_pos", "Z-axis position of photon interaction in plastic scintillator", 200, -25.0, 25.0));
getStatistics().getHisto1D("hit_z_pos")->GetXaxis()->SetTitle("Z-axis position [cm]");
getStatistics().getHisto1D("hit_z_pos")->GetYaxis()->SetTitle("Number of Hits");
  
// Invoking a histogram by title from statistics interface for filling
getStatistics().getHisto1D("hit_z_pos")->Fill(hit.getPosZ())
\end{lstlisting}

\section{Illustrative Examples}
\label{examples}

In this section we present two examples developed with the Framework library.
The first application can be used to perform tests of a prototype PET scanner based on the Monte Carlo simulations of various phantoms.
The simulated scanner was built from a cylindrical layer (radius of 42.5, length of 50 cm) of 384 plastic strips (more details about the MC simulations of the \jpet scanners can be found in~\cite{PMB2018}).
 
\begin{figure}[ht]
\includegraphics[width=0.32\textwidth]{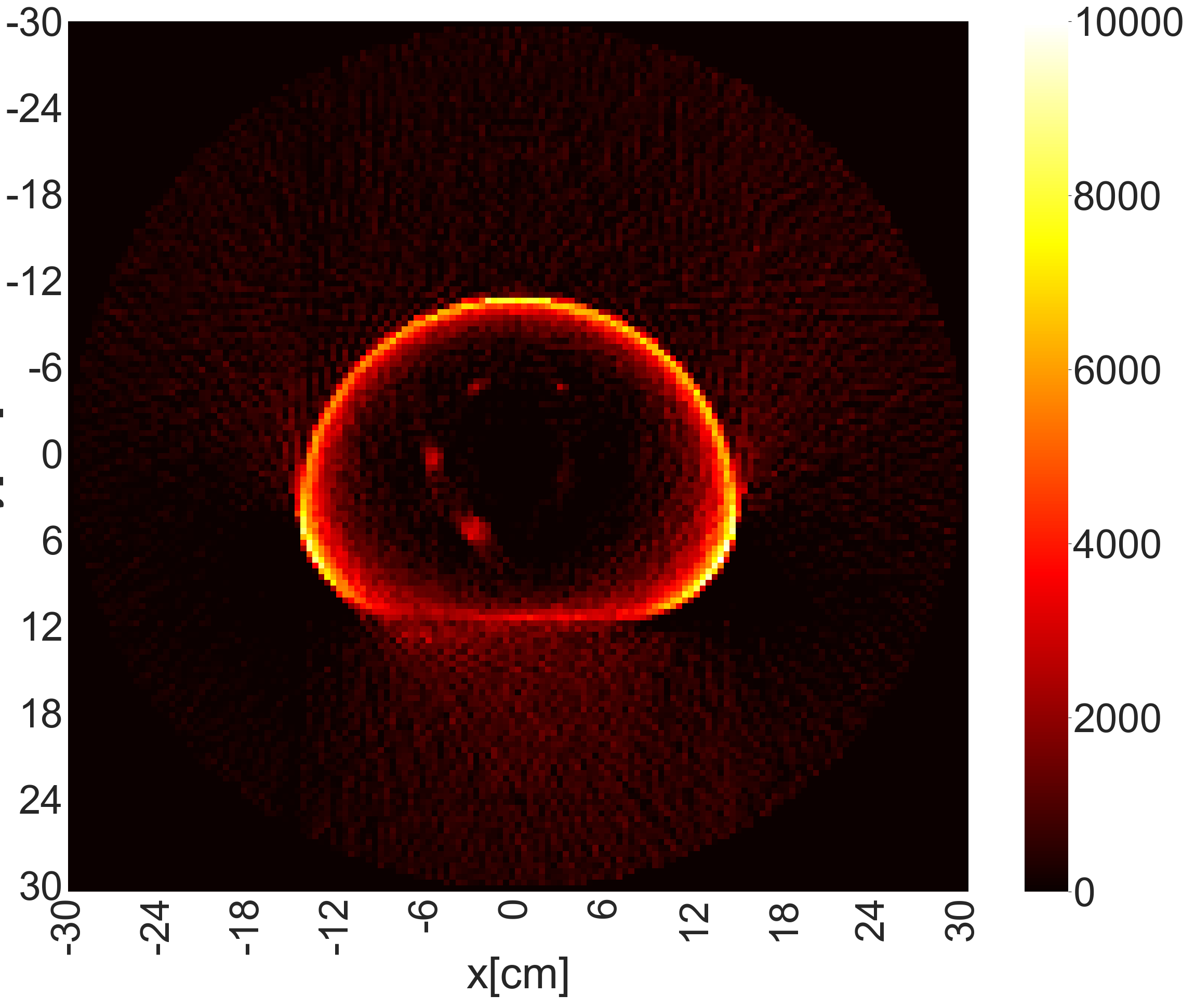}
\includegraphics[width=0.32\textwidth]{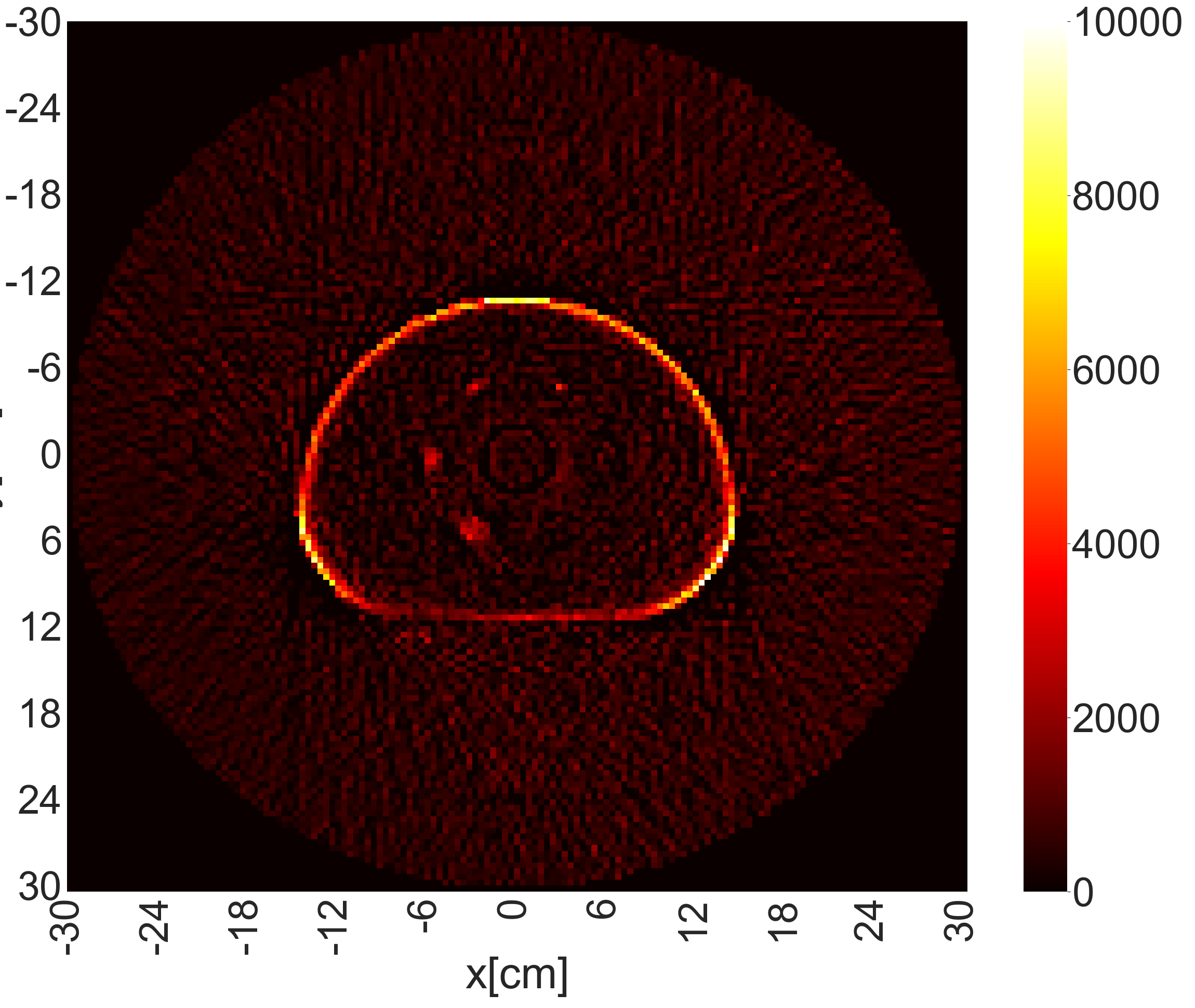}
\includegraphics[width=0.32\textwidth]{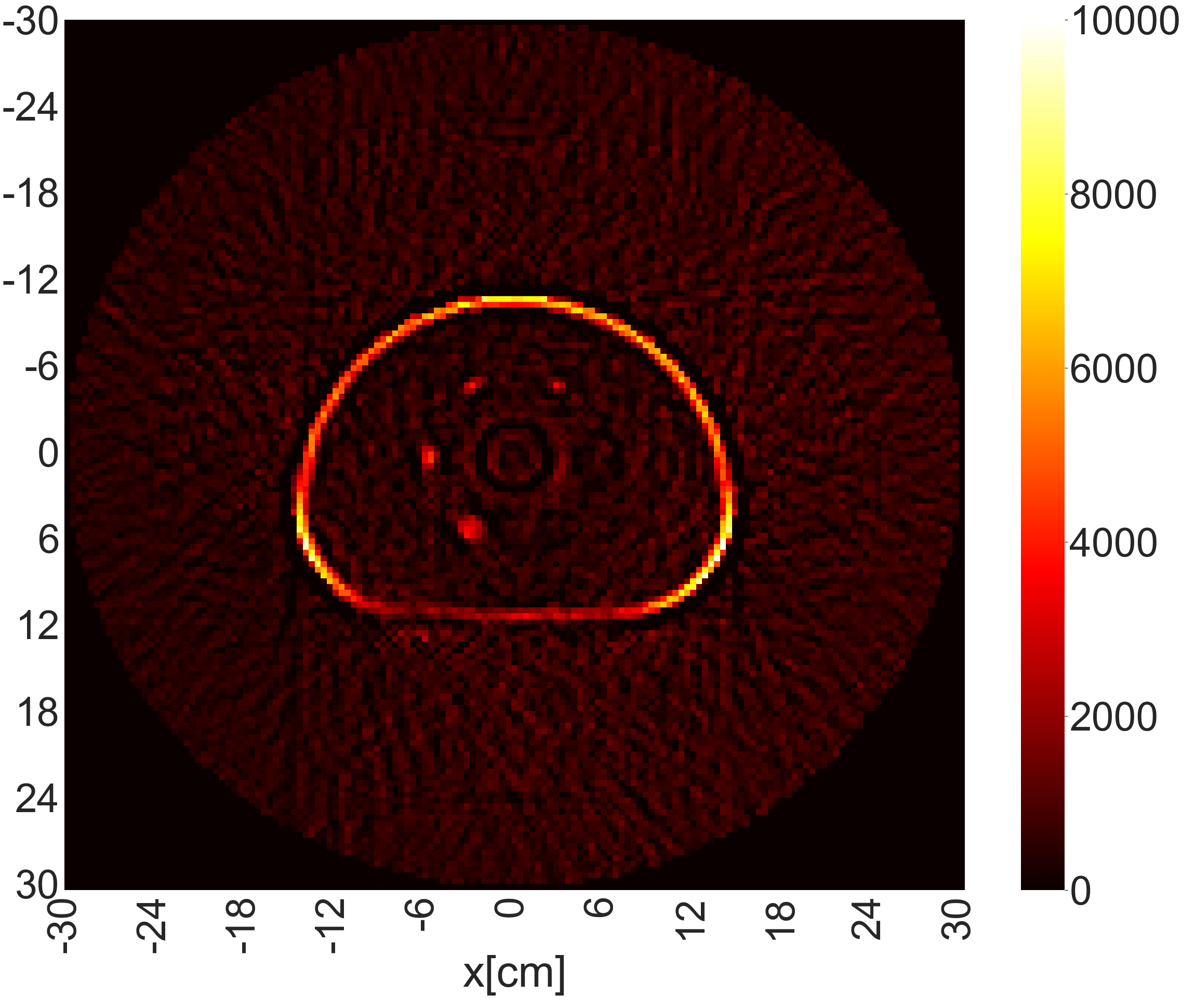}
\centering
\caption{Example image reconstructed with the Time-of-Flight Filtered-Backprojection algorithm with various filters: Ramlak (left), Shepp-Logan (center), Hamming (right). The input sample is based on Monte Carlo simulations of NEMA IEC phantom performed with the GATE package~\cite{Gate_2004},and further processed by the  Framework-based parser which applies the experimental parametrizations to fully imitate a measurement of the scanner.}
\label{fig:toffbp}
\end{figure}

The program loads the data sample generated by the GATE Monte Carlo simulation package and transforms it by smearing the measured observables such as time, energy and position
based on the parametrization of experimental uncertainties determined for a given prototype scanner. This procedures mimics the real measurement effects. Next, the data is reconstructed and finally transformed to a sinogram, which 
serves as an input to the image reconstruction task implementing the Time-of-Flight Filtered-Backprojection algorithm (see Figure~\ref{fig:toffbp}) or can be send to an external image reconstruction package such as STIR~\cite{stir}, CASTOR~\cite{castor} or QETIR~\cite{qetir}. All operations are implemented as consecutive tasks executed by the framework. 

\begin{figure}[ht]
\includegraphics[width=0.33\textwidth]{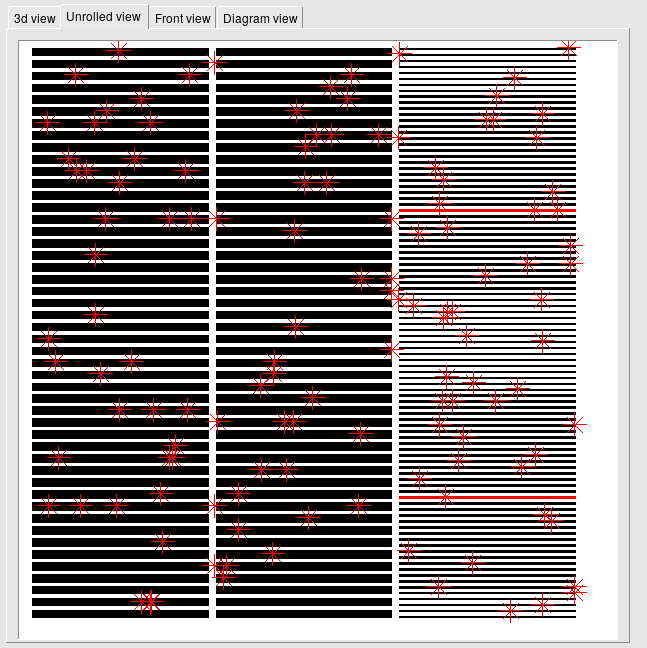}
\includegraphics[width=0.33\textwidth]{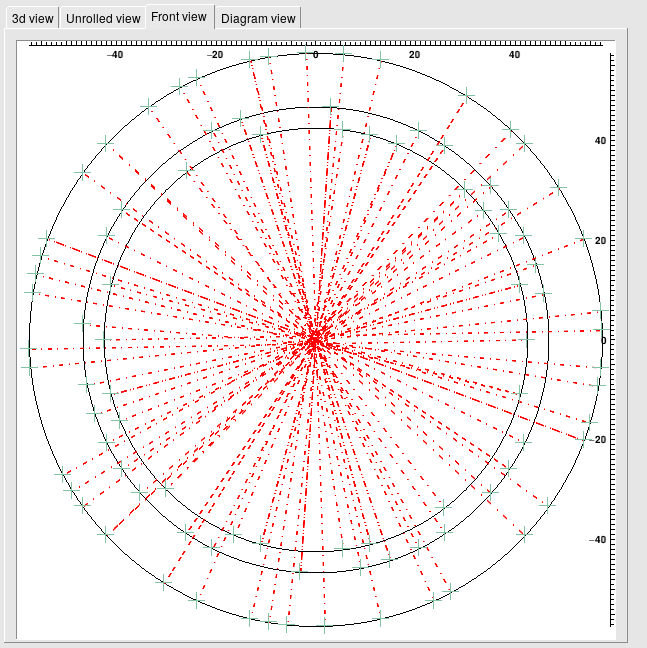}
\includegraphics[width=0.32\textwidth]{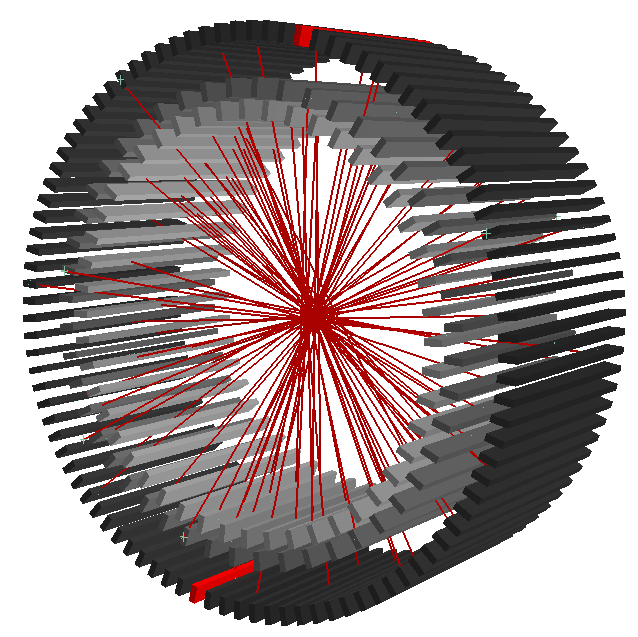}
\centering
\caption{Screenshots of J-PET Event Display \cite{event-display-repo}, (see \ref{sec:event-display}). This example visualizes result of data reconstruction acquired from a test measurement with radioactive source in the center of the scanner. Figures show the model of the J-PET detector, consisting of plastic strips, arranged as 3 concentric cylinders. The right image shows a 3D view, center is a frontal view and left shows these detector elements with positions of reconstructed interactions marked with red stars. In the center and right figures red lines connect pairs of reconstructed positions, that satisfy selection criteria aiming in preparing the sample of events of electron-positron annihilation.}
\label{fig:eventDisplay}
\end{figure}

The second program implements a full reconstruction chain for the real data collected by the 3-layer J-PET scanner. The example reconstruction and analysis is based on the test measurement with the radioactive source placed in the center
of the scanner. The reconstructed results are visualized with the J-PET Event Display tool (see Figure~\ref{fig:eventDisplay}).

\section{Impact}
\label{sec:impact}


Flexibility and robustness of the J-PET Framework library allowed it to be adopted as the main software platform of the J-PET project. The software and applications constructed based on this package have been used for many kinds of scientific studies involving data analyses from the J-PET tomography scanner and will be utilized for future analyses in the fundamental resarch and in the development of various PET scanners prototypes.

\begin{itemize}
\item performance assessment of novel PET scanners ~\cite{Monika-Acta},

\item time calibration techniques for PET scanners~\cite{MS2-time-calib, kd_calib:2020},
\item parametrization of deposited energy in plastic scintillators by Time-over-Threshold measurements~\cite{sushil-tot},
\item implementation of PET image reconstruction techniques such as Kernel Density Estimation, Maximum Likelihood Expectation Maximization ~\cite{MLEM2015} and Time-of-Flight Filtered-Backprojection,
\item development of plastic-based prototype a Positron Emission Mammography scanner~\cite{Shivani:2019mab},
\item studies in positronium annihilation reconstruction and imaging~\cite{alek:pra2016, Jasinska-Moskal2017, NATURE},
\item fundamental research on photon polarization and quantum entanglement~\cite{entanglement2,bHiesmayr},
\item tests of discrete symmetries~\cite{juhi-tsymmetry, alek-tsymmetry},
\item mirror matter searches~\cite{mirror:2019}.

\end{itemize}

The Framework software platform is currently used by scientists from the Jagiellonian University in Kraków, National Centre for Nuclear Research in Warsaw and INFN Laboratori Nazionali di Frascati and has been successfully deployed on different scales starting from laptops and personal PC-s, through mid-size computing clusters to HPC Swierk cluster.

\section{Conclusions}
\label{sec:conclusions}

In this article we presented the features and range of possible applications of the \jpet{} Framework, a C++ based library for data processing and analysis  for PET tomography and for fundamental searches.
The Framework provides tools for the implementation of a wide range
of data reconstruction and calibration procedures
as well as user-level data analyses
and preparation of input for higher-level medical imaging software.
The platform is focused on flexibility in adjusting to dynamically changing prototyping environments and asserting ease of implementation of the required logic by users without programming proficiency while maintaining high processing performance.

Currently, use cases of the \jpet{} Framework span among various data analyses and imaging application of the first \jpet{} device.
In the near future, a new generation light-weight modular \jpet{} scanner with fully digital readout and high mobility will be commissioned along with sibling devices such as a similar-technology-based mammography scanner. The software platform is currently being extended with modules specific to the new devices, which will allow for reusing the higher-level analysis steps with data from new hardware setups.

Despite
having originated solely for the purpose of analysis of data from a single
setup,
the recent expansion of the scope of its usage in the context of \jpet{} demonstrates that 
the flexibility of its architecture allows for use in a wider range of experiments related to nuclear medicine and fundamental studies. 
.

\section{Conflict of Interest}

No conflict of interest exists:
We wish to confirm that there are no known conflicts of interest associated with this publication and there has been no significant financial support for this work that could have influenced its outcome.

\section*{Acknowledgements}


This work was supported in part by the Foundation for Polish Science through the Grant No. TEAM POIR.04.04.00-00-4204/17.
\bibliographystyle{elsarticle-num} 
\bibliography{bibliography}

\end{document}